\documentclass[conference]{IEEEtran}

\usepackage{graphicx}
\usepackage[utf8]{inputenc}
\usepackage{listings}
\usepackage{xcolor}
\definecolor{codegreen}{rgb}{0,0.6,0}
\definecolor{codegray}{rgb}{0.5,0.5,0.5}
\definecolor{codepurple}{rgb}{0.58,0,0.82}
\definecolor{backcolour}{rgb}{0.99,0.99,0.99}

\lstdefinestyle{mystyle}{
    backgroundcolor=\color{backcolour},   
    commentstyle=\color{codegreen},
    keywordstyle=\color{magenta},
    numberstyle=\tiny\color{codegray},
    stringstyle=\color{codepurple},
    basicstyle=\ttfamily\footnotesize,
    breakatwhitespace=false,         
    breaklines=true,                 
    captionpos=b,                    
    keepspaces=true,                 
    numbersep=5pt,                  
    showspaces=false,                
    showstringspaces=false,
    showtabs=false,                  
    tabsize=2
}
\begin{document}

\title{Swim: A Runtime for Distributed Event-Driven Applications}
\author{Chris Sachs, Ajay Govindarajan, Simon Crosby}
\maketitle

\begin{abstract}

Swim\footnote{authors: first@Swim.inc}
extends the actor model to support applications composed of {\tt linked} distributed actors that continuously analyze boundless streams of events from millions of sources, and respond in-sync with the real-world.  

Swim builds a running application from streaming events, creating a distributed dataflow graph of {\tt linked}, stateful, concurrent \emph{streaming actors}  that is overlaid on a mesh of runtime instances.  Streaming actors are vertices in the dataflow graph that concurrently analyze new events and modify their states.  

The Swim runtime streams every actor state change over its {\tt links}  to other (possibly remote) actors using op-based CRDTs that asynchronously update remotely cached actor state replicas.  This frees local actors to compute at any time, using the latest replicas of remote state.  Actors evaluate parametric functions, including geospatial, analytical, and predictive, to discover new relationships and forge or break {\tt links}, dynamically adapting the dataflow graph to model the changing real-world.  

Swim applications  are tiny,  robust and resource efficient, and remain effortlessly in-sync with the real-world, analyzing, learning, and predicting on-the-fly.
\end{abstract}

\section{Introduction}

Streaming events – from users, devices, products, infrastructure, and applications – continue to grow in volume and importance. Events are \emph{state updates} from sources (eg: vehicle location, for a rideshare app), but aren’t \emph{transactions}.   Event-driven applications need to \emph{continuously analyze} boundless event streams, find insights, and respond~\cite{s1}.  Time is ever-present: Events are of ephemeral value, and applications must quickly identify  changes in \emph{system state}  so they can react accurately, and continuously. Applications must concisely capture the past and project into the future, analyzing, learning, and predicting on-the-fly~\cite{k2}.  To enable automation, they must remain \emph{in-sync} with the real-world, delivering responses in milliseconds.
Use cases include automation, click-stream analysis, ad-delivery, VR/AR, online financial services, assembly line automation, IoT and other applications.  

A key infrastructure pattern is \emph{pub-sub}:  Sources \emph{publish} events about \emph{topics} to a \emph{broker}, like Apache Kafka, Apache Pulsar or Redis~\cite{k2,apache,redis}.  Any number of applications can \emph{subscribe} to receive asynchronously delivered events that match their topics of interest.  This frees  application teams from the burden of event stream management - but the broker  does not run applications; it simply queues events by topic.  We want to help developers quickly build applications that \emph{continuously and statefully} consume events to build a dynamic model, and \emph{ continuously and accurately} respond.  
Many brokers present a database abstraction that lets applications  capture events, perform stateless or stateful transformations, create materialized views of streaming data, serve lookups against them, and deliver transformed events back to the broker.  For Apache Kafka this is {\tt ksqlDB}. Analysis is either pull based - a view is evaluated when queried - or push-based - a change to state that affects a view notifies the application to re-evaluate it.   State that is used by many instances is published back to the broker, whereas state used by a single instance is kept in an in-memory database. Time is windowed.
Apache Beam~\cite{apache} is a declarative toolset that helps developers orchestrate processing in sequential stages of a portable pipeline, abstracted from the underlying delivery mechanism, which is  implemented by a \emph{runner} – a “driver” that executes platform-specific commands to manipulate event flow. There are many runners - for raw event streams, brokers, Apache Spark, Apache Flink, databases, and cloud services.  
Beam lets developers focus on the logical composition of pipeline stages. Each is a function that the runtime invokes for each event.  Beam supports parallel operations where appropriate, for example in {\tt map}s.   Transformations can be stateful, but accumulated state is stored by the underlying runner.  As in {\tt ksqlDB}, time is windowed.

Both Swim and Apache Flink build on the actor model. Actors are stateful processes that concurrently compute and then send messages to other actors to inform them of changes. Languages like Erlang~\cite{erlang} and toolkits like Akka~\cite{akka} (used in Flink)  are well known. The key difference between Swim actors and pipeline stages in Beam or push based evaluation in {\tt ksqlDB} is that actors are inherently stateful, atomic and concurrent. Flink actors are functional and state is separately stored.
\begin{figure}[htp]
    \centering
    \includegraphics[width=7cm]{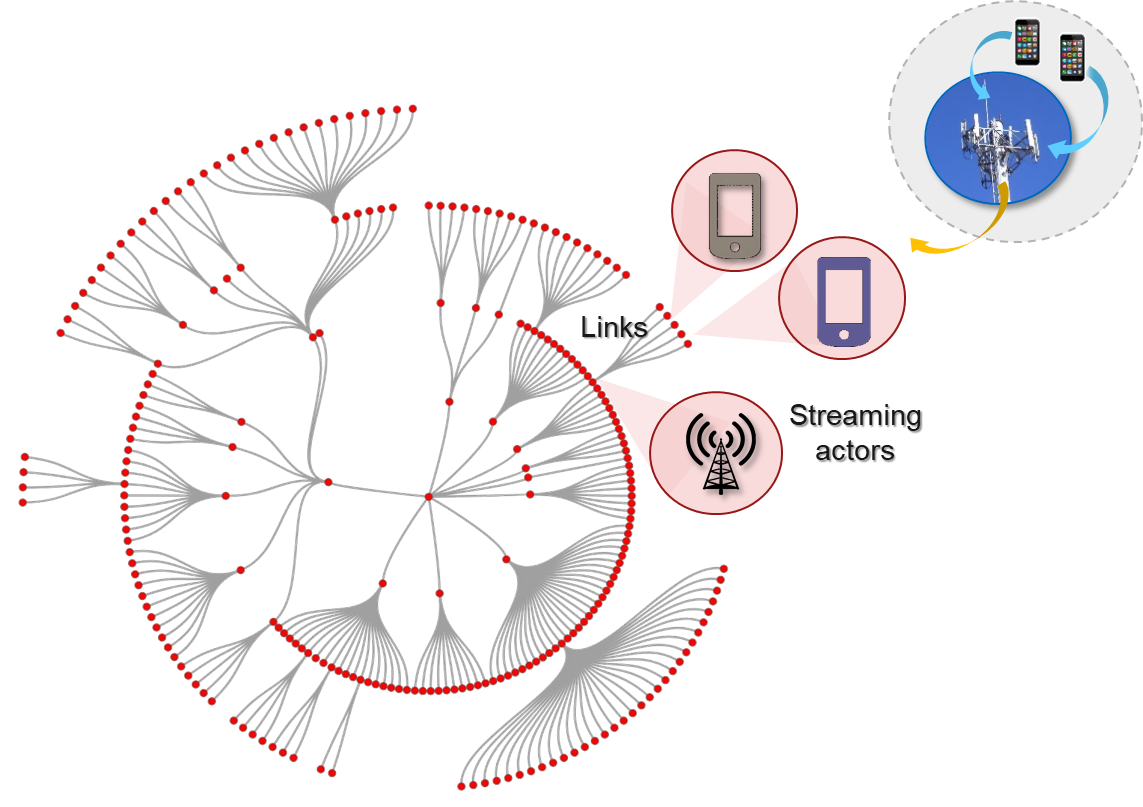}
    \caption{An application is a dataflow graph of {\tt linked} streaming actors}
    \label{fig:dag-mob1}
\end{figure}

Swim is an open-source runtime that extends the actor model to support both \emph{distributed execution} and \emph {streaming}:
\begin{itemize}
\item It builds, runs, and scales \emph{distributed} applications directly from streaming events, in the form of dataflow graphs  of {\tt linked} concurrent \emph{streaming actors}~\cite{act,r16}  (Fig.~\ref{fig:dag-mob1}).  It creates a streaming actor for each unique data source and {\tt links} it as a vertex into  a dataflow graph.     Leaves of the graph are like “digital twins” of event sources, whereas actors that are interior vertices continuously evaluate materialized views or compile indices, so they always have an answer based on the latest data.  Streaming actors compute on new events whenever their inputs change. 
\item Swim \emph{streams} every actor state change over each of its {\tt links}  to other  actors, for use in their analysis. This avoids recipients having to poll, use message passing (and queueing), or manage the overhead of  RPCs.  A {\tt link} is a URI that binds to an actor’s streaming API - extending the \emph{pub-sub}  pattern right through the application.
\item Unlike Apache Flink and Apache Beam that use \emph{statically defined} pipelines, Swim actors fluidly evolve the dataflow graph, making and breaking {\tt links} to represent relationships that they discover by evaluating parametric functions, including geospatial (eg: “near”) and analytical (eg: “correlated”), to track the real-world. 
\item Swim distributes actors in the dataflow graph over a set of (possibly widely) distributed compute \emph{instances}, and periodically relocates actors to balance load.  The runtime orchestrates execution, and manages actor state distribution, persistence, load-balancing, application availability, and recovery from faults.
\end{itemize}

\begin{figure}[htp]
    \centering
    \includegraphics[width=7.5cm]{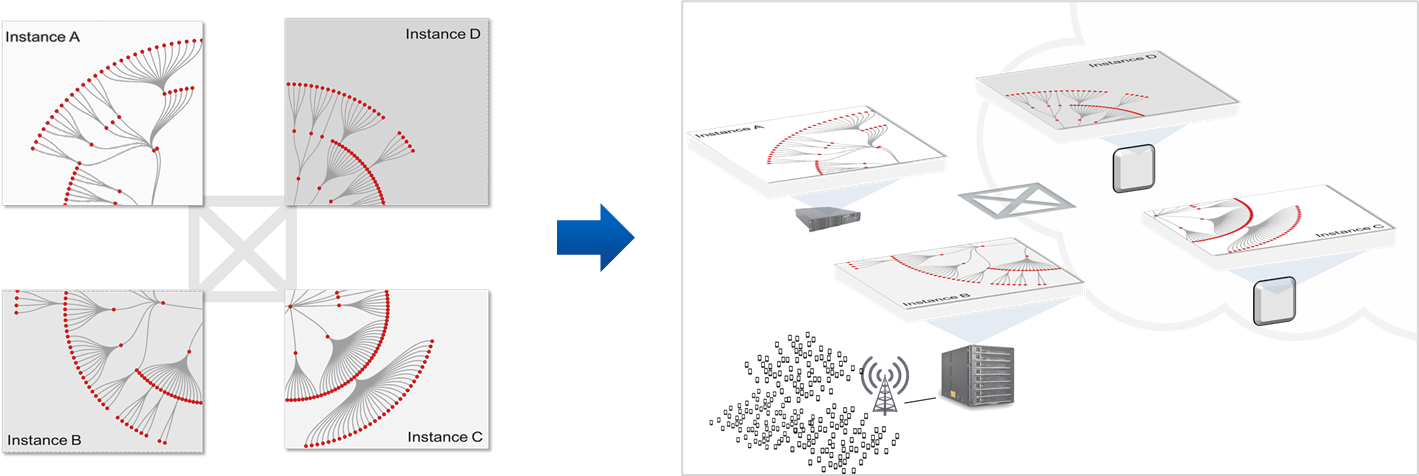}
    \caption{The dataflow graph is distributed over instances}
    \label{fig:deploy}
\end{figure}

For developers, Swim is as easy to use as the single address-space actor model of Erlang~\cite{erlang}, but  instead of requiring a new language, developers use Swim via simple extensions to mainstream languages like Java and Typescript/JavaScript for UIs. Polyglot support is in progress. Applications are tiny, modular, and easy to test, and at runtime they are fast, efficient, robust, and easily scale. 

This paper describes our extension of the actor model to support distributed execution of actor-based dataflow graphs. Graphs that span execution instances raise interesting challenges for actor distribution, data consistency, availability, and response timeliness.  

We present snippets of code from a Swim application that analyzes about 5PB/day (about 15M events/s) to continuously optimize network performance for 240M devices,  and that has  analyzed \emph{exabytes} of data with minimal oversight.  We summarize our experience as “a bit of Dev, but almost no Ops”.

To place our work in context, we focus on three key challenges faced by developers of event-driven applications:
\begin{itemize}
\item Time and timeliness: Many event-driven applications must stay in-sync with the real world,  independent of the scale of the application. Distributed databases struggle to guarantee timely results, but the actor model can help;
\item State management: Today’s “microservice plus database” architectures push complex distributed state management into the database layer, but databases are slow and prevent creative solutions to the consistency problem under partitioning; and
\item Development, deployment and operational management of large distributed applications is complex. 
\end{itemize}
We argue that the actor model, enhanced to support distributed operation using a runtime such as Swim, dramatically simplifies development and continuous operation of event-driven applications.

\subsection{Time and timeliness}

Databases trade off Consistency and Availability under Partitioning~\cite{cap}.  But for event-driven applications CAP seems insufficient. It doesn't include \emph{timeliness} as an attribute of correctness~\cite{k1}:   Yet an event-driven application that uses a distributed database that meticulously keeps replicas in step, but in doing so falls behind the real-world, is a failure.  Many applications need to respond to the \emph{current state} of the system, and don’t need to know \emph{how it got there}  (eg: “where’s my bus \emph{now}?”).  Developers lack the tools to ensure that timeliness is a pervasive concern of an application and its runtime, particularly in a distributed environment.

Partitions are unavoidable, so it is critical to ensure that applications behave correctly under failures. For distributed applications various approaches to state management are used - and unsurprisingly time plays a critical role:  Eventual Consistency (EC) lets database replicas continue to serve users under partitioning, with the promise that “If ... \emph{we wait long enough},  reads will be consistent”~\cite{wait}.  This lacks safety guarantees, but worse, since automation demands immediate responses, “if we wait long enough” is inadequate. If “a system is  EC if, \emph{when all events have been delivered}, all replicas agree ...”~\cite{werner}, then applications that process \emph{boundless} event streams just don’t qualify.  Notably, events are state \emph{updates}, so optimizations for \emph{reads} under partitioning generally aren’t helpful.

Strong Eventual Consistency (SEC)~\cite{bailis} and recent innovations in conflict-free data types offer a path forward: Whereas EC offers only a liveness guarantee, SEC adds the safety guarantee that any two database replicas that received the same (unordered) set of updates will be in the same state. 

Swim offers SEC to {\tt linked} actors to mitigate the effects of delays on distributed applications for which \emph{timeliness} is key:  
\begin{itemize}
\item Swim uses Conflict-free Replicated Data Types (CRDTs)~\cite{lars} to continuously and asynchronously keep distributed \emph{actor state replicas} consistent, decoupling actor execution from state updates. Remote actor states are locally cached in memory so computation at each instance can always proceed without delay, using the latest updates.
\item Under partitioning, Swim strives to ensure \emph{local} responsiveness of surviving instances through careful placement of subgraphs of actors on execution instances that are “near” their data sources, so that the subgraph can remain locally consistent and responsive.
\item Swim actors are persistent, but Swim persists actors only \emph{after} communicating their state changes over their {\tt links} to other actors in the graph.  We call this an “analyze and then store” approach.  Persistence has only one goal: It enables Swim to ensure high availability by deferring operations of unbounded delay (storage), taking them \emph{off the critical path}.  Actors are stateful web services that are always available, and respond in real time.
\item Swim is a vertically integrated stack: Developers focus solely on application logic, and all operational concerns related to running the distributed application are handled by the runtime, with a focus on timely execution and response, accuracy and consistency, and resilience to infrastructure failures.  Moreover, each actor is a stateful web-based microservice whose APIs extend the {\emph pub-sub} metaphor to the individual actor level.  It replaces REST-based polling with a subscription to the actor’s streaming API.
\end{itemize} 
We note that continuous analysis of unbounded event streams is made more complex by timeliness requirements: Analysis is necessarily incremental~\cite{incr,boundless} and many datasets are non-stationary~\cite{taleb}.  Swim has a growing library of tools to help, but there is a rich set of community-led projects to draw from~\cite{apache}. In particular, actors support both windowed and “current state” notions of time.  For many event driven applications event times are vital because the \emph{joint} states of event sources \emph{in time} are key to discovering valuable insights  (eg: “you’re near me \emph{now}”). Applications need to analyze and interpret events given uncertainties of delivery times - but need to keep up first and foremost.

\subsection{State Management}

\begin{figure}[htp]
   \centering
  \includegraphics[width=5cm]{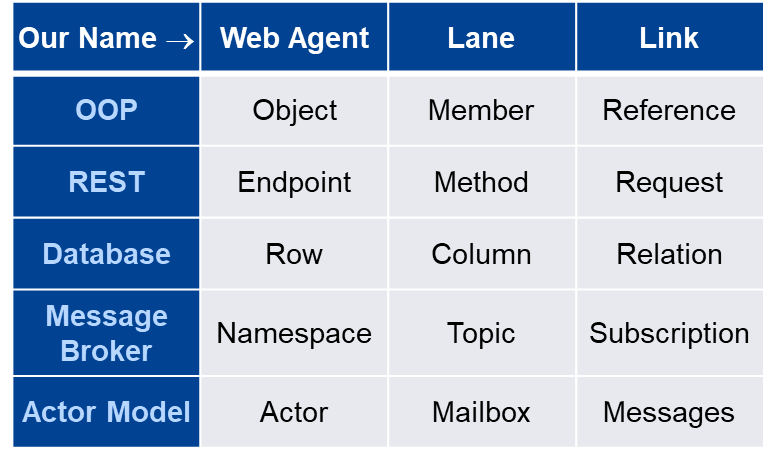}
 \caption{Comparison with other technologies}
  \label{fig:trad-comp}
\end{figure}

How should event-driven applications manage state? Database transactions are familiar for developers, but databases are designed to manage competing \emph{writers}. A good example is online seat booking:  A distributed database must lock a seat (in all replicas) while a booking is completed. Transactions ensure correctness, but at the cost of limiting throughput and increasing latency, which is a problem if correct application behavior demands an immediate response.   That said, handling events as database transactions allows the application front-end to be stateless (thence scalable – like AWS Lambda).  But this pushes state distribution, replication, consistency, and availability management into the database layer, which uses expensive\footnote{Any networked database is at least a factor of $10^6$ slower than the CPU!} protocols to keep distributed replicas in step.  

“Microservice plus database” architectures allow granular code decomposition but make state management harder: Microservices are stateless but relying on a database to store, index or query system state inserts unknown delay into the critical path.  For transactions that \emph{span multiple microservices} Sagas~\cite{sagas} are needed: Each microservice updates part of the database but the entire transaction isn’t complete until the saga completes.  Complex logic is needed to orchestrate computation, and then unwind sagas that fail. Sagas are the opposite of Swim’s actor-based, event-driven model:  Each actor owns and atomically modifies its state, and the runtime streams every state update to all related ({\tt linked}) actors in the dataflow graph, coordinating computation.

\subsection{Simplifying Dev and Ops}

The distributed actor model gives us an opportunity to simplify \emph{“DevOps”} for event-driven applications. We want:
\begin{itemize}
\item {\bf Developers to focus only on application logic.} Today, run-time architecture decisions dictate how applications are written:  Microservices, databases, RPCs, protocols, availability zones (to name but a few) are infrastructure-oriented, operational choices that impact developers, for example by making them deal with distribution, performance, partitioning and recovery.  
\item {\bf Applications that run $24\times 7$.} Today lifecycle management for applications that span clouds and “the edge”, is complex.  A distributed application runtime ought to shield operational teams from the complexities of deployment, distribution, scaling, availability, and persistence.  
\end{itemize}

\subsubsection{Developers}

Microservices have been a powerful addition to the developer toolset. They enable:
\begin{itemize}
\item Granular functional decomposition of application logic, supporting loosely coupled development, testing deployment, and even scaling of functional blocks of code – as services,
\item Simplification: A microservice implements a single function so it can be independently authored (in any language), tested and modified, and then deployed and  scaled in production,
\item Location independence: Microservices often use simple REST APIs  or RPCs, and are designed to be runtime location independent.  They can be bound to specific instances as needed, and capabilities suchas load balancers or service meshes resolve bindings at runtime.
\end{itemize}
But microservices depend on databases for state storage, introducing delay and forcing developers to deal with error conditions.  Although a load balancer can proxy events to stateless microservice instances that are spun up and torn down as needed, the same is not true for the database tier.  To avoid delays and competition for access, database instances need to be carefully sized for their anticipated load.  Moreover, complex materialized views have to be computed when they are used – causing unpredictable delays.  Whilst this may be acceptable for SaaS applications, if the application needs to stay in-sync with the real world, and if its computations (and responses) rely on time (eg:  correlation) the database must continuously evaluate the predicate, which is expensive.

By contrast, the actor model is naturally stateful – state is memory resident in the context of the actor that owns it.  Swim takes this further by making  actors persistent, and by ensuring that remote actor state needed in a distributed application is always cached locally and kept current using a cache coherence protocol.  Further, each actor is a stateful web service that can be accessed  by any authenticated client, using the machinery of the web.
Swim lets developers create and test applications in a single address space and then deploy them unchanged to hundreds of instances.  Actor IDs / addresses are just opaque identifiers that are bound to host-independent URIs.  At runtime time they are bound to the instance where the actor currently resides. 

One might consider the Swim actor model as an implementation of stateful microservices without relying on a database; Swim further augments this by {\tt linking} actors into a relational graph, streaming state changes over their {\tt links}, and using these streamed changes to orchestrate computation.  Streaming avoids the use of a database entirely; instead, stateful actors are memory-resident objects whose streaming APIs enable authenticated entities to access their state changes in real-time as a stream of events.

\subsubsection{Operations}

The goal of the Swim runtime is to make applications resilient and secure, and to orchestrate computation and automate scaling.  Since the reality of event delivery means that applications may need to straddle cloud and edge environments, the runtime must operate in user-mode:
\begin{itemize}
\item Traditional OSes and hypervisors manage individual machine resources, and multi-cloud/edge applications are deployed in VMs or containers~\cite{containers} on an abstraction like Kubernetes~\cite{k8s}.   
\item Moreover, both Swim and its applications may need to span machines with different instruction set architectures, so the runtime and applications need to use an ISA-independent, user-mode virtual machine such as the JVM or LLVM~\cite{llvm}.  
\item The runtime must be able to acquire and release computing resources for the application to enable elastic scaling, if the underlying abstraction (eg: k8s) supports it.
\item It must dynamically distribute computation over the available resources by learning what the application needs, and adapting at in response to load. Specifically:
\begin{itemize}
\item Latency counts: An instance “near” a user (or a robot) can respond in microseconds, in contrast to one on the other side of the planet.
\item Placement counts: Applications span edge-to-cloud:  \emph{Events} are best converted to \emph{state updates} close to their sources, whereas cloud-hosted instances with GPUs are better for \emph{inference}.  
\item Under partitioning, applications must adapt: they should remain responsive locally, and they should repair themselves automatically.  
\end{itemize}
\end{itemize}

Swim shields operational teams from many concerns: Both Swim and its applications run in user-mode on the Graal VM~\cite{graal}.  The platform can also dynamically acquire and release compute instances for an application using an experimental Kubernetes operator.  All runtime instances run the same code and each instance needs only local (or pod) based storage for persistence.  At runtime actors are automatically load-balanced,  relocating them to optimal compute instances.  Applications can be made highly available so they can adapt under failures and automatically recover.   

\section{Architecture Overview}
\label{sec:brief}

\begin{figure}[htp]
    \centering
    \includegraphics[width=7.5cm]{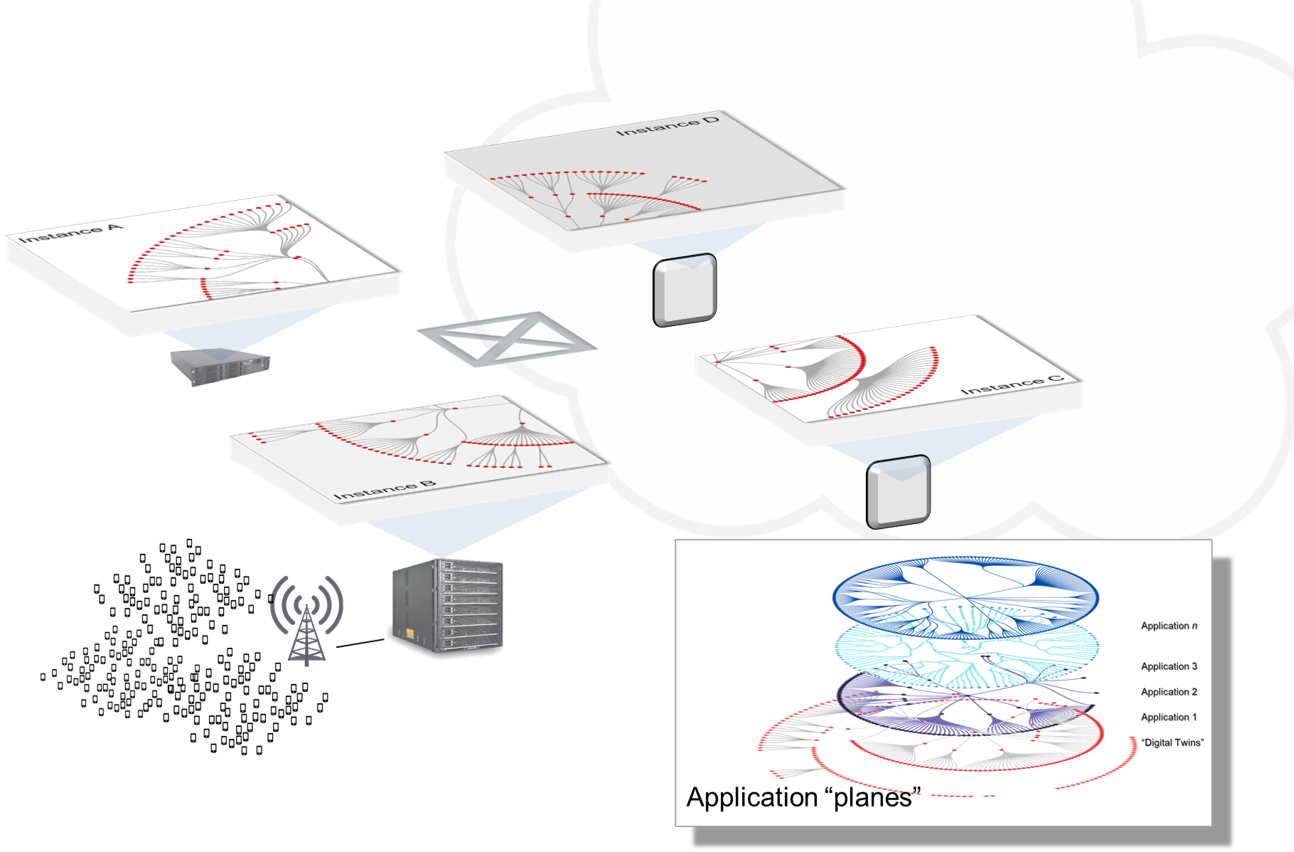}
    \caption{Applications are graphs (\emph{planes}) of Web Agents}
    \label{fig:dag-3d-mob}
\end{figure}

Swim actors are called Web Agents.  They offer innovations in distribution, naming and addressability, statefulness, consistency, and composability.   Figure~\ref{fig:trad-comp} compares Swim abstractions to other familiar technologies.

%\begin{figure}[htp]
 %  \centering
 %   \includegraphics[width=4cm]{gui.png}
  %  \caption{Browser-based JS objects continuously render updates from Web Agents they link to, dynamically %changing on every \emph{click} or \emph{zoom}}
 %   \label{fig:gui}
%\end{figure}

\begin{enumerate}
\item Web Agents are {\bf vertically integrated}: Each is a stateful process \emph{and} also a fully-fledged web service.  Each has a unique URI identifier/address, that can be used like an object reference in a single address space, but in a distributed environment seems just like a REST endpoint. Unlike stateless RESTful microservices that must be polled for updates, however, a Web Agent is stateful, and {\tt linking} to its {\tt lane} is a subscription that continuously synchronizes the state of the {\tt lane} at the {\tt link}er.  
\item Web Agents {\bf compose themselves into a dataflow graph} by {\tt linking}. {\tt Links} are instance-independent URIs, so the application (and the developer) is oblivious to the distribution of Web Agents over instances at runtime. Swim relocates Agents periodically to optimize application performance.
\item A Web Agent can {\bf compute at any time}, using its own state and the \emph{latest state} of every Agent to which it is {\tt linked}, which is cached by the runtime and updated as soon as an update is received, using  pure op-based CRDTs transmitted using a cache coherency protocol, WARP.  
\item Since {\tt links} map to HTTP/2 streams, {\bf Swim UIs can be browser-based}: Each Typescript/JavaScript object that renders a backend Web Agent stays in-sync by {\tt linking} to it, so it receives a continuous stream of updates and can render an always-current view.   {\tt Links} are made and broken on-the-fly, to ensure that the browser only receives updates relevant to the current zoom level on a geo-map.
\end{enumerate}

Below we discuss key architectural choices. We first describe the application model, event processing and state management, then present the developer view.   

\subsection{Application Model}
 
Swim unifies the traditionally disparate roles of database, message broker, job manager, and application server, into a single construct: \emph{Web Agents}, with their {\tt lanes}, and {\tt links}.   

\begin{figure}[htp]
    \centering
    \includegraphics[width=8cm]{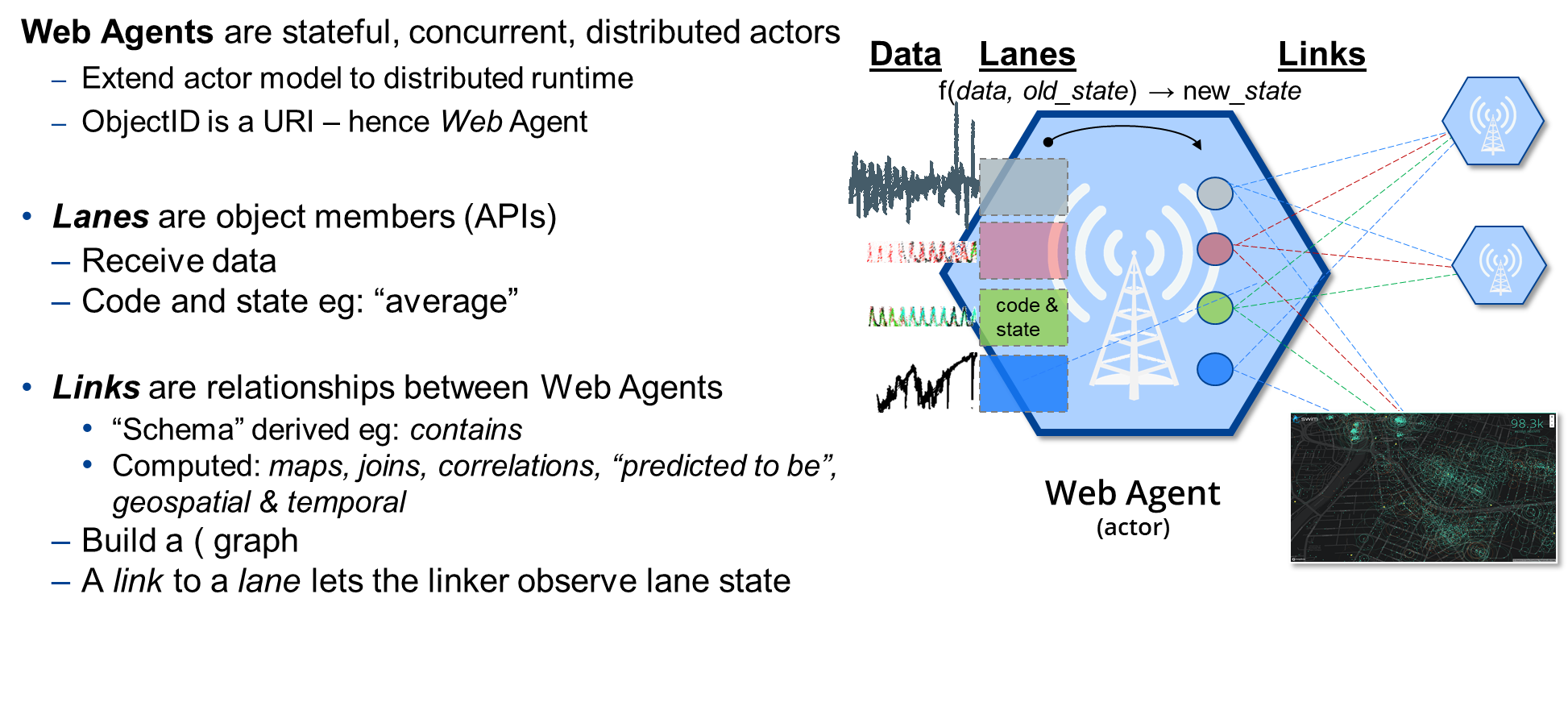}
    \caption{{\tt Lanes} are object \emph{members}. Updates are streamed over {\tt links}}
    \label{fig:dev-model}
\end{figure}

\subsection{Web Agents}
 
Web Agents play one of two roles: 
\begin{enumerate}
\item They are leaves of the dataflow graph that are stateful “digital twins” of event sources that execute OLTP-like logic.  New events are input to a {\tt lane}, which is like an object method. This makes the Web Agent runnable. When it executes it analyzes the new events in the context of its current state and the states of Web Agents it is {\tt linked} to, and atomically modifies its state.
\item They are interior vertices of a graph: A Web Agent re-evaluates its state whenever the state of a {\tt lane} to which it is {\tt linked} changes\footnote{Recalculation can be immediate, periodic, de-bounced etc}, performing  OLAP-like analysis.  A Web Agent can {\tt link} to millions of others. 
\end{enumerate}

A {\tt lane} is a \emph{member} of a Web Agent; it includes code and state.  Agents make and break {\tt links} to {\tt lanes} of other Web Agents as they process events.  {\tt Links} form a fluid, in-memory, cross-instance graph of relationships.   Since Web Agents are processes, they can also interface with traditional applications and infrastructure, for example reading state from, or recording insights in a database.       
 
Swim creates a Web Agent for each unique source in the event stream, and one for each materialized view.  It  distributes Web Agents over execution instances, preferentially placing them in a “local” context – so Web Agents in the same subgraph execute on the same instance, near their event sources.  This eliminates transmission latency since {\tt links} then become local actor references, and it insulates local Web Agents and their subgraphs from the effects of partitioning.  

 \begin{figure}[htp]
    \centering
    \includegraphics[width=7.5cm]{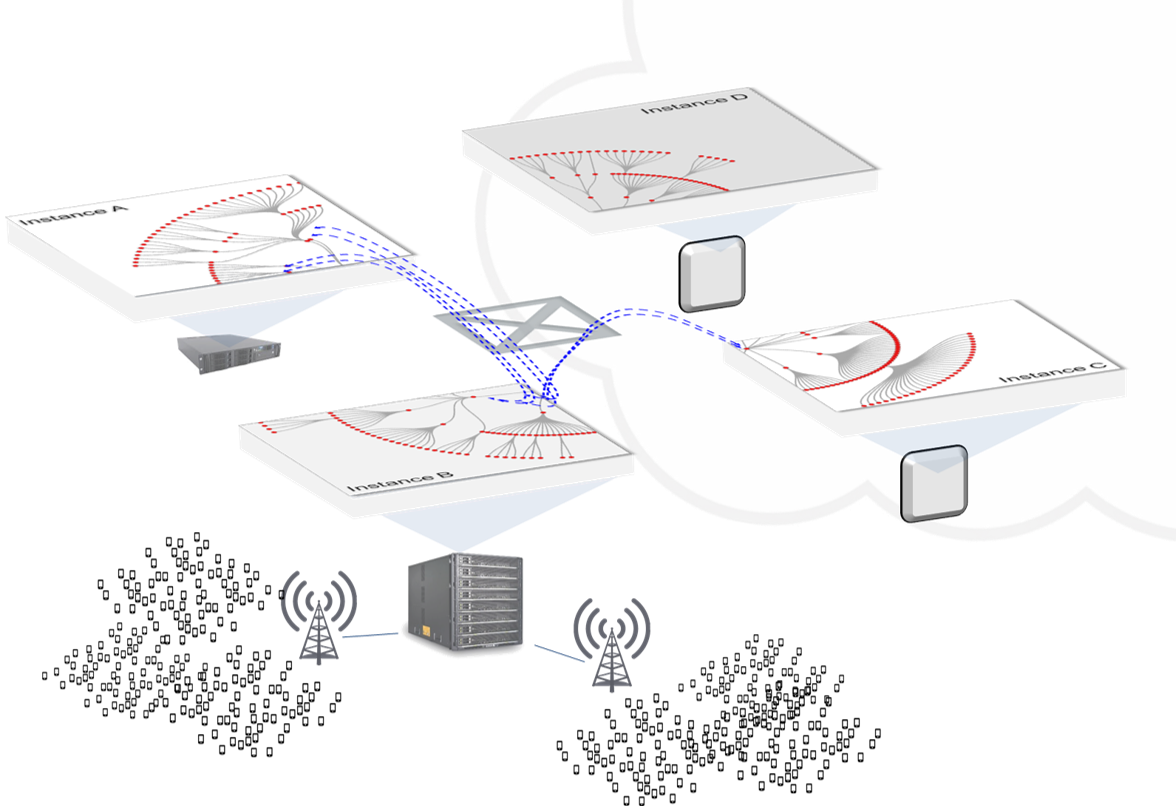}
    \caption{{\tt Links} span instances over HTTP/2 streams (Showing Instance B)}
    \label{fig:dag-3d-mob2}
\end{figure}

Loss of an instance means that the application dataflow graph loses some of its vertices -  the Web Agents that were running on the failed instance. If an application needs it, Swim delivers high availability by replicating each runtime instance, and ensuring that every remote {\tt link} held by Web Agents on the standby instance is read-only.  The standby receives all updates (and its Web Agents compute on them) but it does not stream its Web Agent state changes to remote Web Agents.  Instead, it behaves like an application client (eg: a UI) that simply observes state changes. In the event of failure of the primary instance, the secondary can take over in an instant using an election algorithm like RAFT~\cite{raft}.  

Under network partitioning that severs the graph, the effects depend on Web Agent distribution over instances.  This offers an opportunity to implement a Web Agent type-specific reaction to faults (see Fig.~\ref{fig:dag-instances2a-fault}). Section~\ref{sec:fault} discusses partitioning, availability, and timeliness in detail.

\subsubsection{Event Sourcing}
\begin{figure}[htp]
    \centering
    \includegraphics[width=7.5cm]{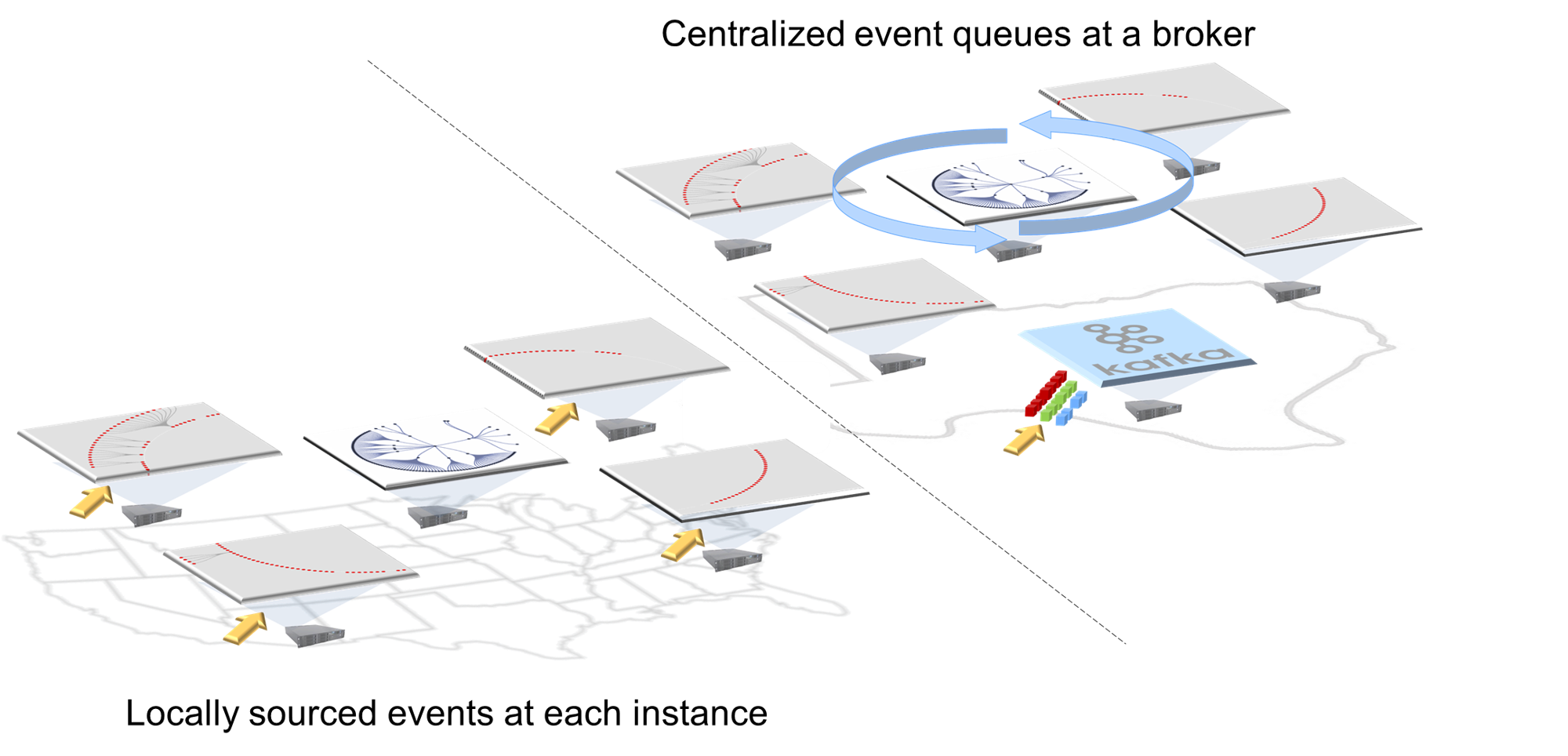}
    \caption{Events can be sourced locally at each instance, or centrally at a broker}
    \label{fig:event-sourcing}
\end{figure}

Any Swim instance can receive exogenous events and distribute them to the appropriate “digital twin” Web Agent anywhere in the application, even if the Agent is running on a remote instance.  This is particularly useful if the application  fuses two or more streams in real-time, where events are sourced from different brokers or using different protocols.  There are two typical patterns for event processing (Fig.~\ref{fig:event-sourcing}):
\begin{enumerate}
\item In a widely distributed application, events will be delivered locally to a “nearby” instance to minimize latency. Careful placement of Web Agents and their subgraphs also ensures local analysis.
\item Some organizations centralize their event streams at a broker like Apache Kafka~\cite{k1}; Swim creates and distributes Web Agents across the instances in a cluster, initially placing Agents on instances using a hash ring. 
\end{enumerate}
Web Agents  are periodically load-balanced and may be moved dynamically to a runtime instance that best suits their CPU, memory, and GPU needs.  Swim preferentially places subgraphs on the same instance to minimize inter-instance state transmission.  Swim can dynamically acquire and release compute instances for an application through an experimental Kubernetes operator~\cite{k8s}; new instances join the mesh and Web Agents are load-balanced to them.
 
Whereas classical actors use message passing in a single address space, Swim applications need a  simple programmatic naming/addressing abstraction that \emph{hides distribution} of Web Agents from the developer, a mechanism that enables Web Agents to seamlessly and immediately \emph{share their state updates} while ensuring coherence and state consistency, and a way to \emph{orchestrate distributed processing}.  The latter two are part of the runtime - addressed in section~\ref{sec:core}.
 
\subsection{{\tt Link}s}
 
A {\tt link} provides a way for Web Agent $a$ to subscribe to a {\tt lane} offered by Web Agent $b$, whether $b$ is in the same address space or on a remote node.  When  $b$ {\tt links} to a {\tt lane} offered by $a$, the runtime records the link; whenever $a$ changes its state, the runtime streams the changed state from $a$ to $b$, but $b$ is unaware of this: It simply reads the state (locally, from an in-memory cached replica of $a$’s state).  For developers, a {\tt link} is a reference to a {\tt lane}, dynamically resolved by a routing proxy.  It resolves to a local Web Agent reference and member if the target is in the same address space, and a locally cached replica of $b$’s state if it is remote.  {\tt Links} can be “observe-only” too: Browser-based clients that render remote Web Agents can {\tt link} to their {\tt lanes} to receive a continuous stream of state updates to render a real-time UI without polling.

\subsection{{\tt Lane}s}
 
Each {\tt lane} is identified by a URI that uniquely identifies the Agent and its API.  {\tt Lanes} are typed with associated semantics. These let the runtime take on data intensive operations that are found, for example, in database query engines.  All deliver events on changes.
\begin{itemize}
\item {\tt Value Lanes}  hold properties of Web Agents.
\item {\tt Map Lanes} hold collection properties of a Web Agent, and are consistent for updates and deletes.
\item {\tt Join Value Lanes} hold properties for {\tt joining} value lanes.
\item {\tt Join Map Lanes} hold collections of properties between Web Agents.
\item {\tt Demand Value Lanes} hold properties of Web Agents.
\item {\tt Demand Map Lanes} consistently hold collection properties of Web Agents.
\end{itemize}

\section{Developer View}
\label{sec:app}

Swim hides the complexity of actor creation, distribution, execution, load-balancing, scaling, and resilience, giving developers a simple, application-focused abstraction that is more akin to object-oriented programming. 

\begin{figure}[htp]
    \centering
    \includegraphics[width=6.5cm]{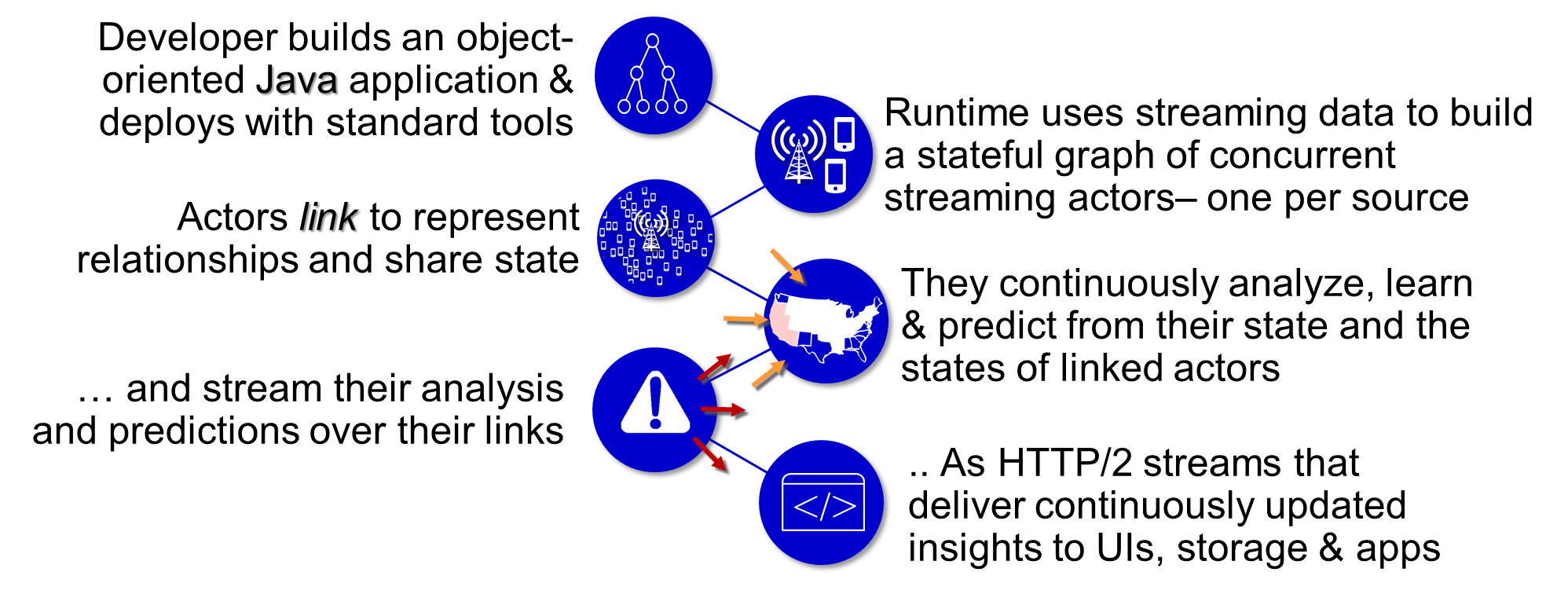}
    \caption{The Swim developer view}
    \label{fig:dev-flow-mob}
\end{figure}
    
A Java developer creates a Web Agent by simply extending the {\tt AbstractAgent} class.  Below we summarize our key choices:
\begin{itemize}
\item  	Together, Web Agents, {\tt lanes}, and {\tt links} implement a continuous consistency model that is transparent to developers.  
\item  	Web Agents are stateful, concurrent objects, so developers don’t need to worry about databases or  actor-relational mapping, or any of the challenges related to distribution (eg: RPCs), failure or recovery. Agents are internally persistent. (On every update the runtime writes the Web Agent state to a local log used to reconstruct state in the event of a crash. This is a built-in capability optimized for performance and reliability.  Notably, state is persisted \emph{after} updates have been streamed to {\tt linked} actors.)
\item  	The only way to modify the state of a Web Agent is through a {\tt link} to one of its {\tt lanes}.  The Agent controls the exposure of its internal state. 
\item  	Though as many Web Agents execute in parallel as there are CPU cores, they are atomic without locks.
\item  	A Web Agent reference is a host-independent URI.  Decoupling an Agent’s logical address from that of its host  instance makes a {\tt link} to it independent of its current runtime instance, so it can be relocated as needed.  This comes with the (tiny) overhead of requiring a lookup on reference.
\end{itemize}
 
The example below is drawn from an open-source application that continuously scores the connection quality for mobile devices, cell towers and the network of a large mobile provider.  The application continuously predicts the connection quality for each device and optimizes network configuration  for the anticipated load. The application is a few thousand lines of Java, but at runtime it services over 5M  events/s and analyzes about 5PB of events per day on 40 instances, distributed over 25 regional data centers.  A simplified version of the application is available as an online demo.  The code snippets are intended to give the reader a feel for the familiar object-oriented style of Java programming that Swim permits.  Specific Swim constructs are hidden in Java extensions.

\begin{lstlisting}[language=java]
  /*  agent for the ENodeB function */
public class eNodeBAgent extends AbstractAgent {
  @SwimLane("status")  ValueLane<Value> status;
  @SwimLane("kpis")  ValueLane<Value> kpis;
  @SwimLane("RANLatest")   /*  New data */
     ValueLane<Value> RANLatest = this.<Value>valueLane()
         .didSet(this::didSetRANLatest);
  @SwimLane("RANHistory")
     MapLane<Long, Value> RANHistory = this.<Long, Value>mapLane()
         .didUpdate(this::didUpdateRANHistory);

  /*  REST endpoint to poll ENodeB */
  HTTPResponse<?> onRequestSummary(HttpRequest<Value> request) {
    final Value payload = this.status.get().concat(this.kpis.get());  
    /*...*/
    return HTTPResponse.from(HttpStatus.OK).content(entity);
  }

/* An Agent for a geo-region */
public class RegionAgent extends AbstractAgent {
  @SwimLane("status")   ValueLane<Value> status;
  @SwimLane("geometry") ValueLane<Value> geometry;
  /* KV map for live status */
  @SwimLane("subRegions") JoinValueLane<Value, Value> subRegions;
  /*...*/
}
\end{lstlisting}
 
A {\tt link} between two Web Agents triggers continuous state replication between their runtime instances.  Either of a pair of {\tt linked} Web Agents  can update its  shared state consistently. A {\tt link} has a {\tt downlink} and an {\tt uplink}. The {\tt downlink}  is held by the endpoint that opened the {\tt link} and the {\tt uplink}   by the endpoint that received the {\tt link} request.
 
To open a {\tt link}, a developer creates a {\tt downlink}, and specifies the address of the target Web Agent and the name of the {\tt lane} (the URI) they want to {\tt link} to:
\begin{lstlisting}[language=java]
 public void RANLatency() {
   if (latencyLink == null) {
     latencyLink = downlink()
         .hostUri(REGION_URI)
         .nodeUri(Uri.from(nodeUri().path()))
         .laneUri("latency")
         .onEvent(this::didSetRemoteLatency)
         .open();
   }
 }
\end{lstlisting}
 
{\tt Downlinks} allow developers to access a {\tt lane}'s state.  If the target is remote, then the {\tt downlink} uses the locally cached version of the remote state without blocking or waiting.  Developers can also set the state of a lane through a {\tt downlink}.
\begin{lstlisting}[language=java]
 void didSetRemoteLatency(Value newValue) {
   latency.set(newValue);
 }
\end{lstlisting}
An observer can observe state changes to remote {\tt lanes} by registering callback functions on its {\tt downlinks}.
\begin{lstlisting}[language=java]
latency.didSet(name => console.log(`connected to ${name}`));
\end{lstlisting}
When no longer interested in the state of a remote lane, an application can close a {\tt downlink} to stop receiving updates.
\begin{lstlisting}[language=java]
latency.close();
\end{lstlisting}

\section{Swim Core}
\label{sec:core}

The Swim runtime is responsible for deploying applications across distributed instances, building and scaling the application layer graph of Web Agents, load balancing instances, ensuring that Agents have the resources they need, persisting the state of the running application, transmission of actor state changes over their links, responding to partitions, and recovery from failures.

\begin{figure}[htp]
    \centering
    \includegraphics[width=7.5cm]{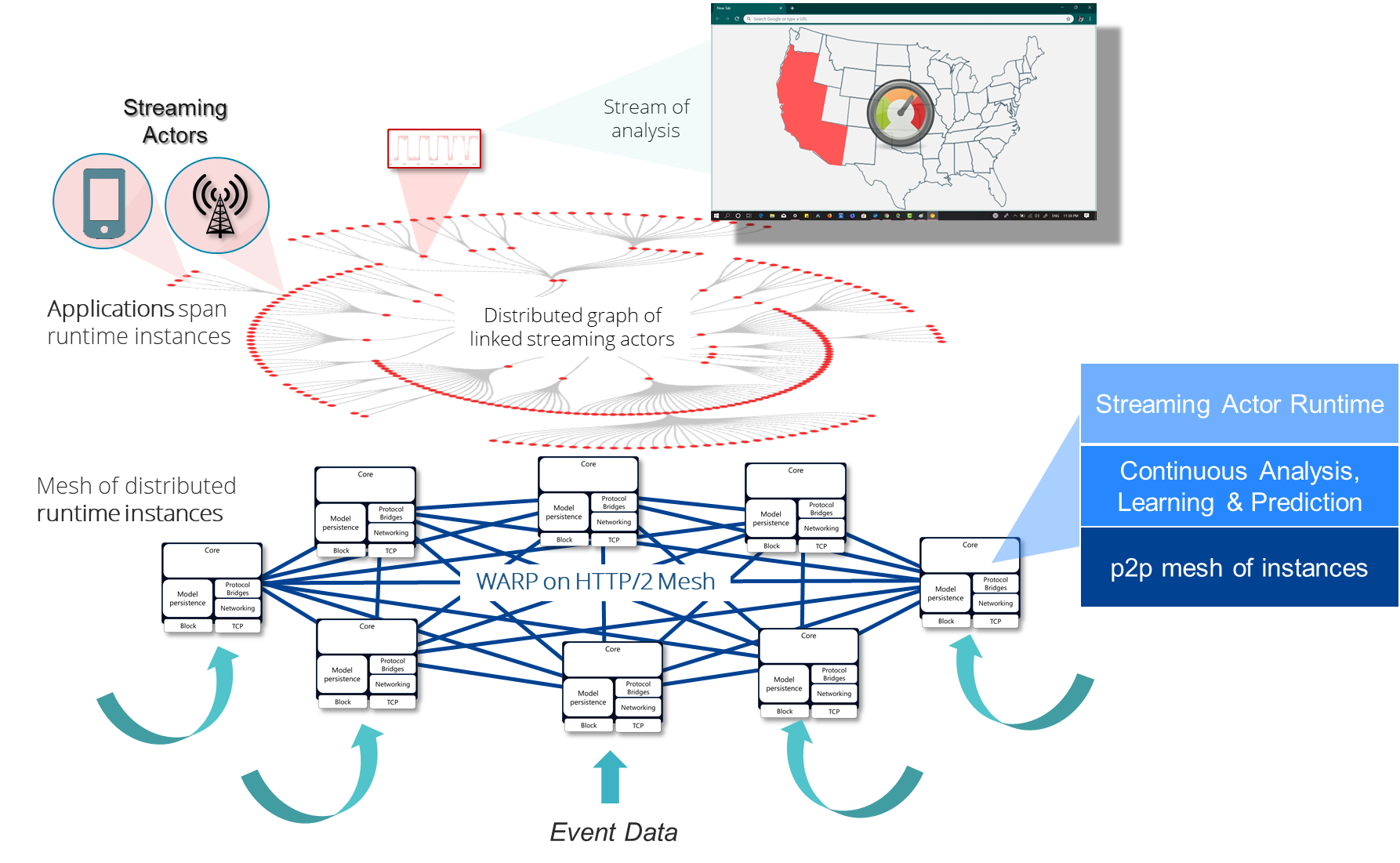}
    \caption{Instances connect in a mesh}
   \label{fig:app-dag2-mob}
\end{figure}

We focus below on the capabilities that enable  the platform to build, run, scale and repair applications while ensuring that they can respond in a \emph{timely manner}. 

\subsection{Actor Streams}
\label{sec:http2}
All communication between instances is managed by the runtime. Developers are unaware of the fact that an application may be distributed.  Web Agent state is in memory, so timely computation depends on Agent-to-Agent communication.  Swim adopts and extends the concepts of rate-limited, prioritized reactive streams~\cite{reactive} to ensure continuous consistency for Web Agent state across distributed instances.  

\begin{figure}[htp]
    \centering
    \includegraphics[width=6cm]{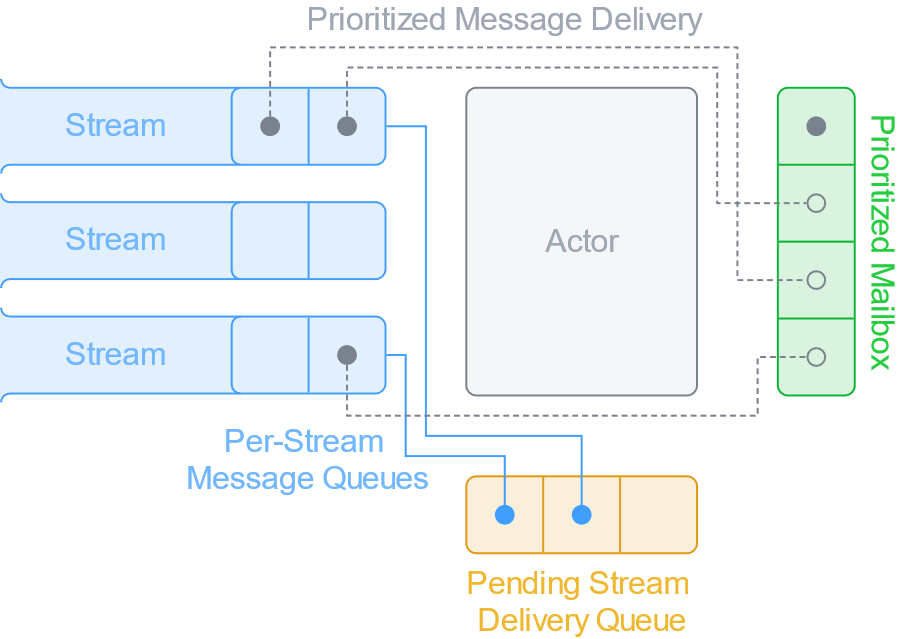}
    \caption{Schematic diagram of  actor streams}
    \label{fig:Web Agent-streams}
\end{figure}
 
An \emph{actor stream} is bound to each Web Agent endpoint of a {\tt link}.   It maintains its own bounded queues for sending and receiving events.  Communication is back-pressure regulated to maintain timeliness, prevent buffer bloat, and to prevent high-rate streams from starving high priority, low-rate streams.    
 
\begin{figure}[htp]
   \centering
    \includegraphics[width=5.5cm]{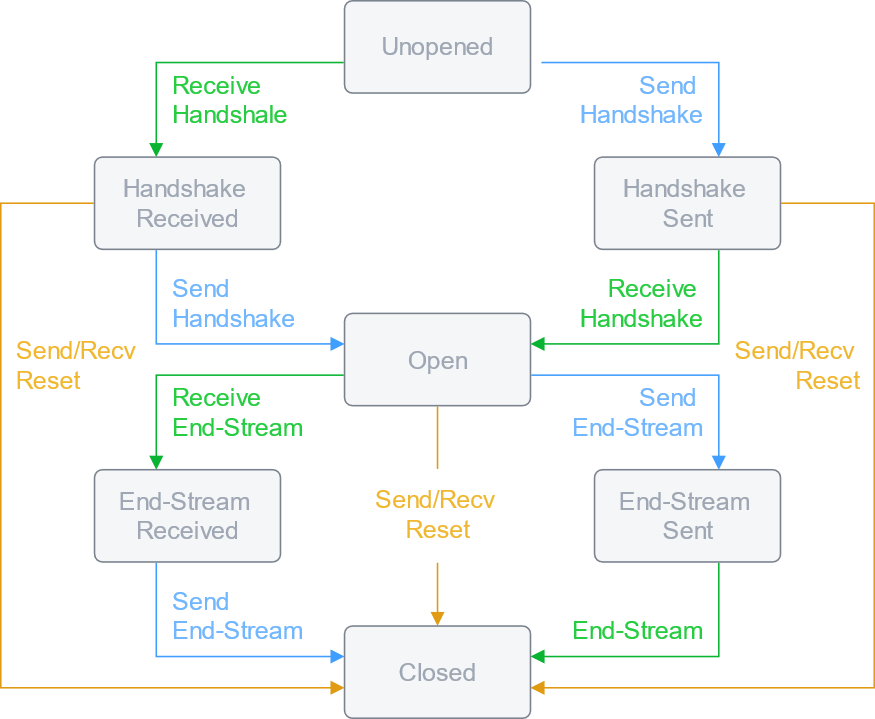}
    \caption{Stream lifecycle}
    \label{fig:stream-lifecycle}
\end{figure}
 
When Web Agent $a$ {\tt links} to $b$:
\begin{itemize}
\item A new actor stream is registered with $b$ by inserting a handshake event into the stream's send queue, appending the stream to $b$’s delivery queue, and scheduling $b$ to run.
\item When $b$ runs, it dequeues the new stream, responds to the enqueued handshake, and stores a reference to the stream. 
\item An actor stream can be forcibly reset by atomically setting a reset flag and enqueueing the stream in both Web Agents’ delivery queues. Streams are reset if an inter-instance connection fails.
\end{itemize}

Each Web Agent has a pending \emph{stream delivery queue} for  streams with events waiting.  This decouples the flow  of events into streams from the processing of streams.  It ensures fairness and preserves timeliness: A Web Agent dequeues a stream, then dequeues events from that stream.  A separate delivery queue is used for streams with a priority above a certain threshold. This ensures that events above a certain priority always take precedence.  
 
\subsection{Event Delivery}

\begin{figure}[htp]
    \centering
    \includegraphics[width=7.5cm]{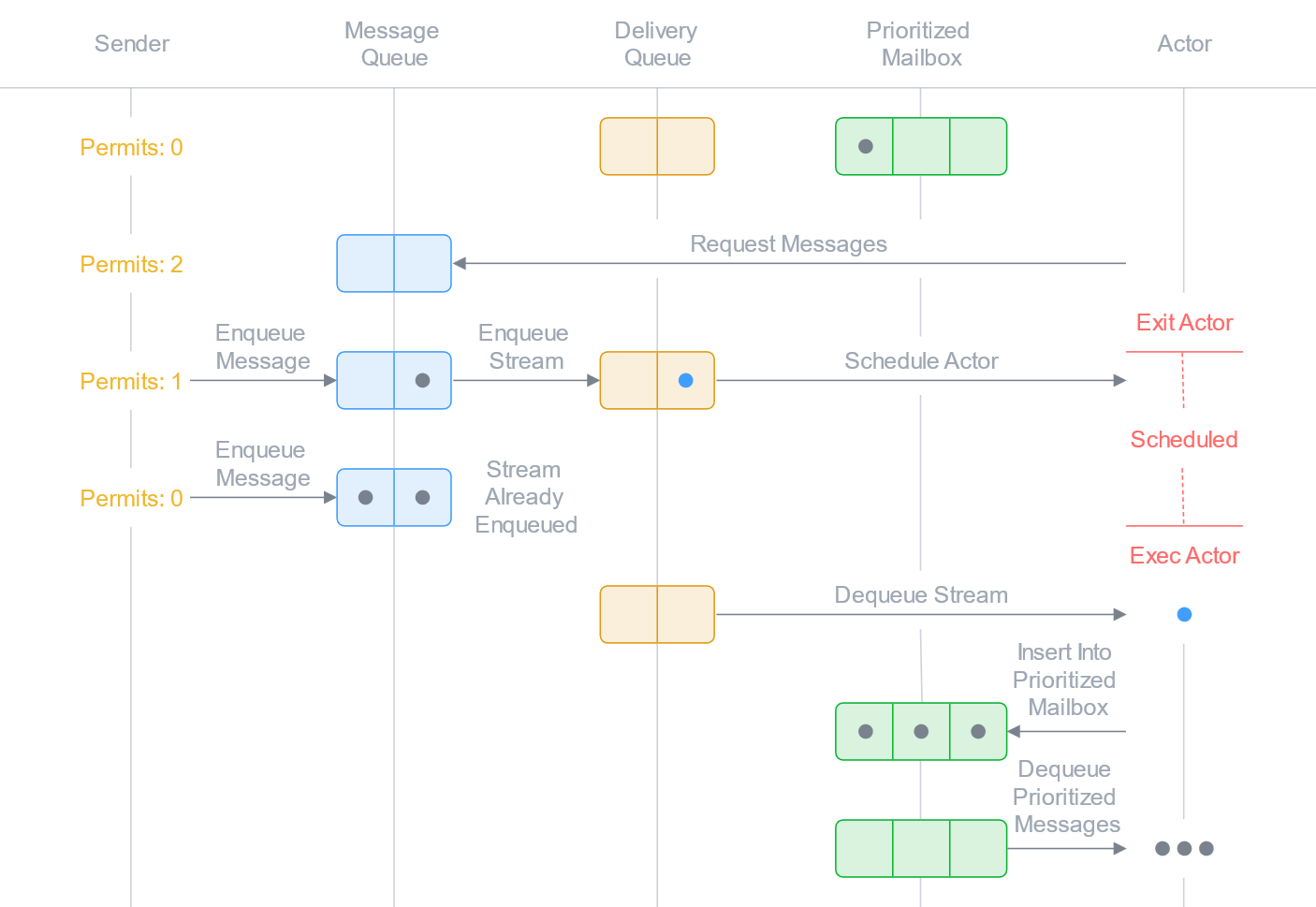}
    \caption{Event delivery}
    \label{fig:event-delivery}
\end{figure}

Enqueueing an event into an actor stream is non-blocking and concurrent. An event  can only be enqueued  when the sender has a permit. (Backpressure is fundamental for end-to-end timeliness). When an actor stream has at least one event queued, a reference is enqueued in the Web Agent's stream delivery queue, and the Web Agent is scheduled to run.  When the  Web Agent runs, it dequeues streams from its delivery queue, and repeatedly dequeues pending events from the  stream, and inserts them into  its mailbox subject to a per-stream limit and a mailbox size limit.

\subsection{State Synchronization}
 
Swim uses Conflict-free Replicated Data Types (CRDTs) ~\cite{bartosz,lars,quilt}, to asynchronously stream every Web Agent state change over its {\tt links}.  Remote instances cache state for locally executing Web Agents with {\tt links} to remote Agents. Asynchronous CRDT update by the runtime decouples state updates from application processing.  We are the first (as far as we know) to use CRDTs to asynchronously update locally cached \emph{actor states}.
 
\begin{figure}[htp]
    \centering
    \includegraphics[width=7cm]{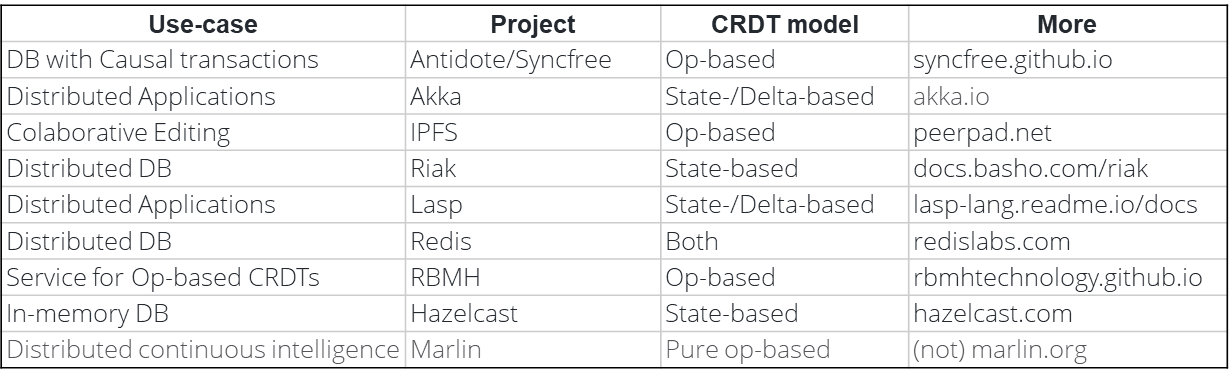}
    \caption{CRDT use in major projects}
    \label{fig:crdts}
\end{figure}
 
CRDTs emerged in the last decade to permit efficient SEC in distributed databases (Fig.~\ref{fig:crdts}).   We want  to permit distributed, concurrent updates to cached local replicas of remote Web Agent state, and to resolve inconsistencies after the fact by merging replicas. CRDTs  enable us to resolve concurrent updates without conflicts~\cite{r7,r8}. There are two types that offer SEC: 
\begin{itemize}
\item State-based CRDTs~\cite{r2} support data structures for which changes are commutative. They use a gossip protocol  to disseminate the entire state of each CRDT to each replica, on every change. States are merged  by a function that is commutative, associative, and idempotent. The merge is a join for any pair of states. Delta state CRDTs send only changes~\cite{r3}.
\item Operation-based CRDTs (op-based) send update \emph{operations} rather than the entire state. They are more concise, but operations cannot be dropped or duplicated, and must be delivered in causal order.  Pure op-based CRDTs reduce the metadata size and are well suited to Web Agent state dissemination.
\end{itemize}
 
\begin{figure}[htp]
    \centering
    \includegraphics[width=5.5cm]{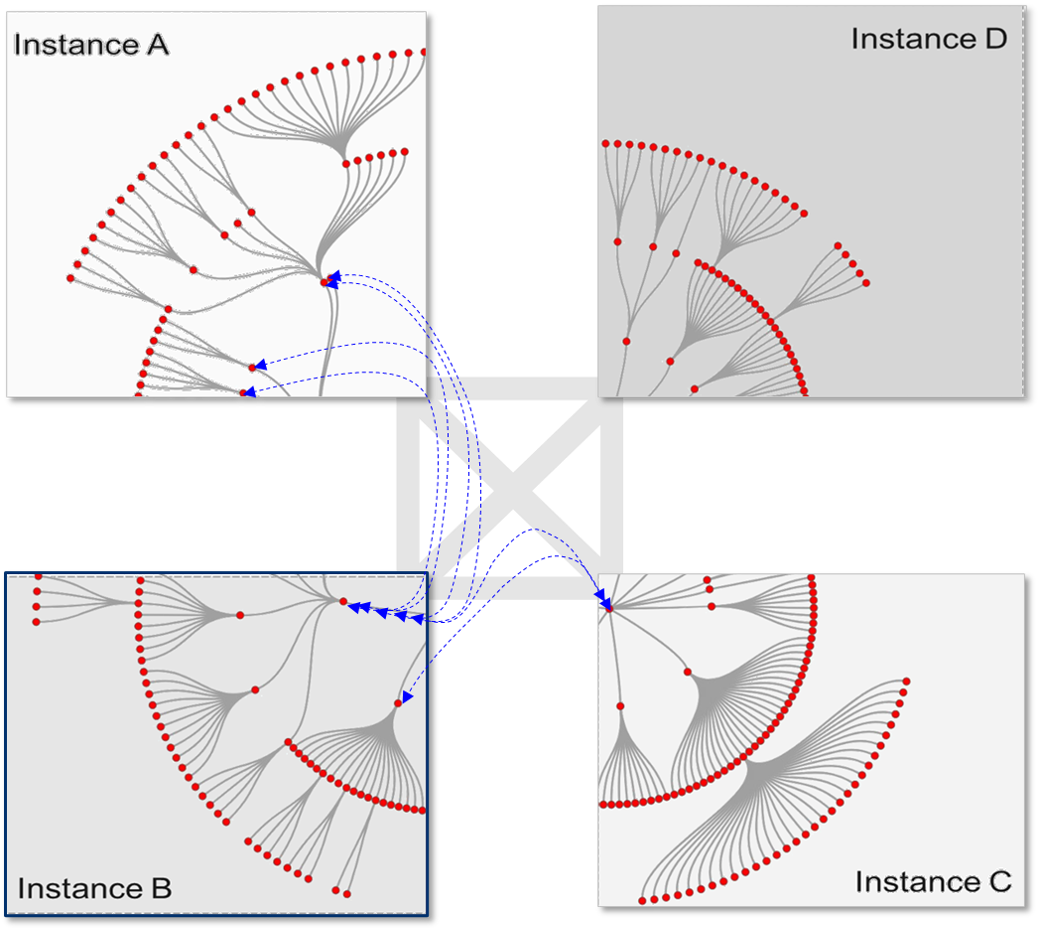}
    \caption{Instances cache replica state for remote {\tt linked} Web Agents.  Showing {\tt links} at $B$}
    \label{fig:only-cache}
\end{figure}
 
Whereas a  distributed database must keep \emph{every replica} consistent, Swim streams a Web Agent’s state changes only to other Web Agents to which it is {\tt linked} (Fig.~\ref{fig:only-cache}).   An instance only needs to keep replicas for  remote Web Agents to which its local Agents are linked.  The  runtime manages CRDT updates, freeing the application layer to compute at any time.  Op-based CRDTs encode the update \emph{operations} to be made to cached state, avoiding the need to make remote updates visible at the application layer (in contrast to RPCs that make the operation explicit to the developer).
   
\subsubsection{WARP}
The Web Agent Remote Protocol (WARP) synchronizes bidirectional  {\tt lanes} between Web Agents, delivering CRDTs over HTTP/2.   WARP endpoints make what appear to be RPC calls over HTTP/2 streams, but the semantics are non-blocking and are used for delivery only.  WARP implements a cache coherency protocol that keeps the distributed application instances in-sync.  It enables Swim to offer continuous consistency.
 
WARP supports server and client state replication for {\tt linked} Web Agents. Server-side state replication is used for peer-to-peer state synchronization between back-end application instances Client replication occurs when a client (eg: a browser UI) synchronizes with a given server, and the server mediates replication between the client and Web Agents on other instances in the cluster. Client replication hides the topology of the cluster, masks internal replication metadata, and prevents clients from polluting the cluster with state replication events.
 
Both server and client replicas can be read-only, write-only, read-write, or observe-only. Read-write replicas do not need to wait for propagation to read-only or write-only members before committing events. Read-only replicas cannot influence the causal history of the CRDT, and write-only replicas cannot promise consistency. Data types with commutative operations can be implemented as pure op-based CRDTs using standard reliable causal delivery, but data types with non-commutative operations use a PO-Log, a partially ordered log of operations.  

WARP is a Tagged Causal Stable Broadcast (TCSB)~\cite{bartosz} protocol that provides causality information on event delivery and informs the sender when delivered events become causally stable, allowing  PO-Log compaction. When the local  replica of a remote Web Agent is updated, Web Agents that are linked to the remote Web Agent are made runnable so they can process the changed remote state.  If a remote Web Agent becomes inaccessible due to partitioning, the CRDT is marked as stale – effectively “use with caution”.  If the connection between two instances is broken, all {\tt links} to Web Agents on the remote instance are reset.
 
\subsection{{\tt Links} map to HTTP/2 Streams}
HTTP has evolved into a general purpose multiplexed streaming protocol. With WARP on HTTP/2, the job of routing a CRDT update to a remote instance reduces to the need to deterministically route a proxied HTTP/2 stream for the relevant Web Agent, within the distributed application instances that are interconnected in a mesh. An HTTP/2 stream between runtime instances is  a full-duplex communication channel between {\tt linked} Web Agents.  
 
The use of HTTP/2 also puts each Web Agent on the web (most deployments are on private Intranets), and makes its {\tt lanes} accessible to any authenticated HTTP/2 client.  This also allows browser based UIs to monitor CRDT updates to render remote Web Agent state in real-time.   A {\tt lane} reference reduces to a REST API in the case that the caller makes a single invocation, and the API simply returns the current state. 

\subsection{Availability under Partitioning}
\label{sec:fault}
 
\begin{figure}[htp]
    \centering
    \includegraphics[width=7cm]{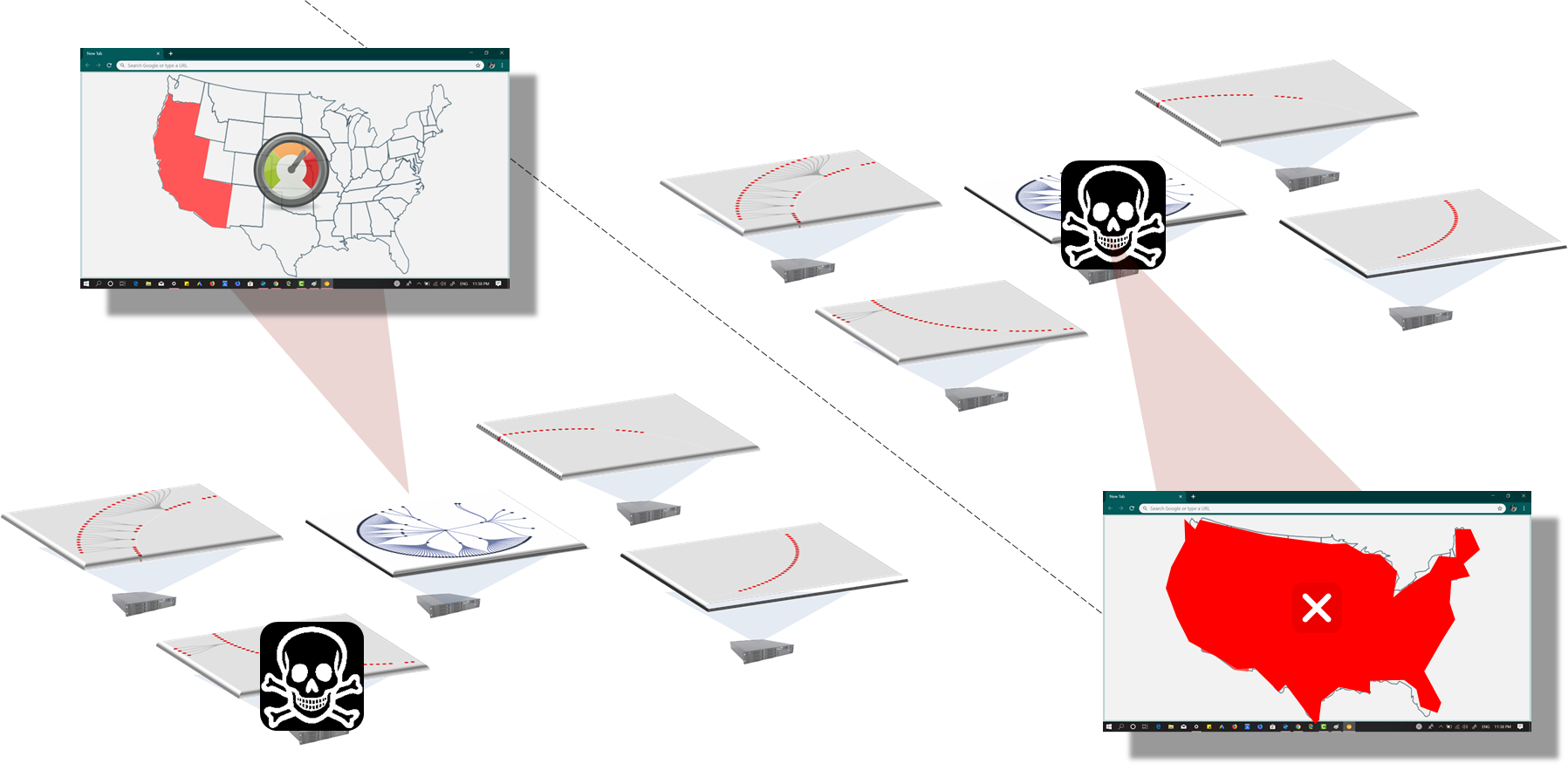}
    \caption{Mapping Web Agents to instances determines the effects of faults}
   \label{fig:placement-and-failure}
\end{figure}

For event-driven applications, new models of consistency and availability are needed that respect timeliness~\cite{k1}, given the inevitability of partitions.  Agent placement strategies determine the effects of faults (Fig.~\ref{fig:placement-and-failure}).  Loss of part of the dataflow graph might lead to a regional outage but might not impact other regions.  Loss of an instance that hosts continuously evaluated materialized views could be critical.

Swim implements a model that we term \emph{continuous consistency} - Web Agent states and their replicas are kept continuously updated subject only to network delays.  Actor (and replica) state is memory-resident, but changes are continuously and lazily persisted locally using an append-only log  for fast writes; log compaction is also lazy and continuous.  Web Agents can adapt to partitioning in a type-specific way and respond with information that can help users.  As shown in Fig.~\ref{fig:dag-instances2a-fault}, if a remote instance is unresponsive (the HTTP/2 connection to its instance breaks); replicas of its Web Agents’ states will become \emph{stale} and {\tt links} will be reset.  At instances with no {\tt links} to Web Agents on the partitioned instance, computation can proceed unperturbed (Instance D).  

On failure and subsequent recovery of an instance, the log allows the runtime to recover its recent state, and that of local application Web Agents. It rebuilds its connections to other instances, local Web Agents are restarted with their recent states, and if the Web Agents are leaves of the application dataflow graph they resume external event processing.  Actors that are materialized views will automatically receive the current state of all Web Agents on their {\tt links} (just like a new {\tt link} and can immediately resume processing using the current state. Election of the primary instance in a HA configuration uses RAFT~\cite{raft}.

\subsubsection{Future Work}
Swim’s distributed actor model offers powerful advantages for \emph{continuous operation} of applications at scale, but it does not address the CI/CD approach to \emph{continuous  deployment} of new code.  Whereas stateless microservices that each implement a single function can be upgraded easily, changing the behavior of a stateful Web Agent type is not possible.  Deployment of new functionality for a specific Web Agent type can only be achieved by causing each instance to restart with the new code.  Our current work aims to achieve this using Web Assembly: Each Web Agent will be a WASM executable that can be independently upgraded in a sandbox.
We are developing polyglot language support and the Swim core is being rewritten in Rust.  

\section{Conclusion}

In distributed applications at scale, it is not feasible to “store then analyze”  events. Automation and an increasing need for real-time insights and responses demands that applications analyze, learn, and predict on-the-fly.  
Parametric relationships between sources call for continuous evaluation using a dynamic  model of the \emph{system} which is simplified using an actor-based approach. Actor-based applications can assemble themselves from events, and are easy to test, deploy and operate. In our experience, distributed actor-based applications are easy to develop, fast and efficient, scale well and are robust.  

Our approach offers a powerful alternative to traditional database-centric application design that offers users enormous efficiency and performance benefits, and developers a simple object-based model that can seamlessly scale to billions of actors at runtime.  Swim offers the research community an open-source platform to investigate continuous consistency and timeliness for event-driven applications, without sacrificing availability.
 
%%%%%%%%%%%%%%%%%%%%%%%%%%%%%%%%%%%%%%%%%%%%%%%%%%%%

\end{document}